\documentclass[submission,copyright,creativecommons]{eptcs}
\providecommand{\event}{LFMTP 2018} 
\usepackage{breakurl}             
\usepackage{underscore}           

\usepackage[bw]{agda}
\usepackage{stmaryrd}
\usepackage{textgreek}
\usepackage{catchfilebetweentags}
\usepackage{newunicodechar}
\usepackage{mathtools}
\usepackage{amsmath}
\usepackage{amsthm}
\usepackage{amsfonts}
\usepackage{amssymb}
\usepackage{stackrel}

\DeclareUnicodeCharacter{411}{\ensuremath{\lambdabar}}
\DeclareUnicodeCharacter{8799}{\ensuremath{{\stackrel{?}{=}}}}
\DeclareUnicodeCharacter{8788}{\ensuremath{{\coloneqq}}}
\DeclareUnicodeCharacter{8348}{\ensuremath{{_t}}}
\DeclareUnicodeCharacter{8345}{\ensuremath{{_n}}}
\DeclareUnicodeCharacter{65288}{\ensuremath{{(}}}
\DeclareUnicodeCharacter{65289}{\ensuremath{{)}}}
\DeclareUnicodeCharacter{8336}{\ensuremath{{_a}}}
\DeclareUnicodeCharacter{8759}{\ensuremath{\dblcolon}}
\DeclareUnicodeCharacter{8718}{\ensuremath{\blacksquare}}
\DeclareUnicodeCharacter{9632}{\ensuremath{\blacksquare}}
\DeclareUnicodeCharacter{8667}{\ensuremath{\Rightarrow}}
\DeclareUnicodeCharacter{955}{\ensuremath{\lambda\!\!}}
\DeclareUnicodeCharacter{956}{\ensuremath{\mu}}
\DeclareUnicodeCharacter{545}{\ensuremath{\rightarrow}}
\DeclareUnicodeCharacter{945}{\ensuremath{\alpha}}

\title{Formalisation in Constructive Type Theory of  Barendregt's Variable Convention for Generic Structures with Binders}
\author{Ernesto Copello
    \institute{Department of Computer Science \\ The University of Iowa, USA}
    \email{ernesto-copello@uiowa.edu}
\and
Nora Szasz \qquad\qquad \'Alvaro Tasistro
    \institute{Facultad de Ingeniería\\Universidad ORT Uruguay}
    \email{szasz@ort.edu.uy \qquad\qquad tasistro@ort.edu.uy}
}

\def\titlerunning{Formalisation of  Barendregt's Variable Convention for Generic Structures with Binders}
\def\authorrunning{E. Copello, N. Szasz, A. Tasistro}

\begin{document}

\newcommand{\AgdaInductiveConstructorFoot}[1]{{\small\AgdaInductiveConstructor{#1}}}
\newcommand{\AgdaDatatypeFoot}[1]{{\small\AgdaDatatype{#1}}}
\newcommand{\AgdaFunctionFoot}[1]{{\small\AgdaFunction{#1}}}
\newcommand{\Nat}{\ensuremath{\mathbb{N}}}
\newcommand{\Real}{\ensuremath{\mathbb{R}}}
\newcommand{\Var}{\ensuremath{\mathsf{V}}}
\newcommand{\Vars}{\Var}
\newcommand{\Terms}{\ensuremath{\mathsf{\Lambda}}}
\newcommand{\Subst}{\ensuremath{\mathsf{\Sigma}}}
\newcommand{\Restr}{\rest}

\newcommand{\ddefs}{\ensuremath{=_{\mathit{def}}}} 
\newcommand{\ddef}[2]{\ensuremath{#1 \ddefs #2}} 

\newcommand{\free}{\textsl{free}}
\newcommand{\freer}[2]{\ensuremath{#1 *#2}}
\newcommand{\samefree}[2]{\ensuremath{#1 \sim_{*} #2}}
\newcommand{\ssamefree}[3]{\ensuremath{\res{\samefree{#1}{#2}}{#3}}}
\newcommand{\fvst}[1]{\ensuremath{\mathit{FV}\!#1}}
\newcommand{\comm}[1]{\textsf{#1}}
\newcommand{\cnst}[1]{\textsl{#1}}
\newcommand{\vart}{\textsf{V}}
\newcommand{\codom}{range}
\newcommand{\res}[2]{\ensuremath{#1\downharpoonright#2}}
\newcommand{\rest}{\textsf{P}}
\newcommand{\reseqs}{\ensuremath{\cong}}
\newcommand{\freshr}[2]{\ensuremath{#1\ \#\,#2}}
\newcommand{\fresh}[3]{\ensuremath{#1\ \#\,\,\res{#2}{#3}}}
\newcommand{\subseq}[3]{\ensuremath{\res{#1 \reseqs #2}{#3}}}
\newcommand{\sexteq}[2]{\ensuremath{#1 \reseqs #2}}
\newcommand{\alpeqst}{\ensuremath{\sim_{\alpha}}}
\newcommand{\alcnv}[2]{\ensuremath{#1\alpeqst #2}}
\newcommand{\subscnv}[3]{\ensuremath{\res{#1\alpeqst #2}{#3}}}
\newcommand{\longred}[2]{\ensuremath{#1\,\mbox{$\to\!\!\!\!\!\to$}\,#2}}
\newcommand{\idsubst}{\ensuremath{\iota}}
\newcommand{\upd}[3]{\ensuremath{#1, #2 {:=} #3}}
\newcommand{\subsap}[2]{\ensuremath{#1 #2}}
\newcommand{\choicest}[1]{\ensuremath{\chi\,#1}}
\newcommand{\scomp}[2]{\ensuremath{#1 \circ #2}}
\newcommand{\empc}{\ensuremath{\cdot}}
\newcommand{\ccons}[2]{\ensuremath{#1 , #2}}
\newcommand{\indom}[2]{\ensuremath{#1 \in\!\mathsf{dom}\,#2}}
\newcommand{\lkup}[2]{\ensuremath{#1\,#2}}
\newcommand{\delst}[2]{\ensuremath{#1 - #2}}
\newcommand{\tj}[2]{\ensuremath{#1 \vdash #2}}
\newcommand{\incls}{\ensuremath{\preccurlyeq}}
\newcommand{\incl}[2]{\ensuremath{#1 \incls #2}}

\newcommand{\preds}{\ensuremath{\rightrightarrows}}
\newcommand{\pred}[2]{\ensuremath{#1 \preds #2}}
\newcommand{\predsz}{\ensuremath{\rightrightarrows_{0}}}
\newcommand{\predz}[2]{\ensuremath{#1 \predsz #2}}

\newcommand{\bred}[2]{\ensuremath{#1 \breds #2}}
\newcommand{\boreds}{\ensuremath{\rightarrow}}
\newcommand{\bored}[2]{\ensuremath{#1 \boreds #2}}
\newcommand{\rtc}[1]{\ensuremath{#1^{*}}}

\newcommand{\epf}{\hfill\ensuremath{\Box}}
\newcommand{\epff}{\ensuremath{\Box}}

\newcommand{\del}[1]{\st{#1}}
\newcommand{\re}[2]{\st{#1}\hl{#2}}
\newcommand{\add}[1]{\hl{#1}}
\newcommand{\betaalpha}{\ensuremath{\rightarrow_\alpha}}
\newcommand{\betaaster}{\ensuremath{\rightarrow_\beta^*}}
\newcommand{\betaarr}{\ensuremath{\rightarrow_\beta}}
\newcommand{\betacontr}[2]{\ensuremath{#1 \triangleright #2}}
\newcommand{\type}[3]{\ensuremath{#1 \vdash #2 : #3}}
\newcommand{\first}[2]{\ensuremath{\text{First}\ #1\ #2}}
\newcommand{\cL}{{\cal L}}
\newcommand{\breds}{\ensuremath{\twoheadrightarrow}}
\newcommand{\p}{\ensuremath{\rightrightarrows}}
\newcommand{\pp}{\ensuremath{\rightrightarrows_T}}
\newcommand{\betar}{\ensuremath{\rightarrow_\beta}}
\newcommand{\conc}{\ensuremath{+\!\!+}}
\newcommand{\rbet}{\ensuremath{\rightarrow_\beta}}

\newcommand{\pnbeta}{\ensuremath{\pn\!\!\beta}}
\newcommand{\betarn}{\ensuremath{\rightarrow_{\beta n}}}
\newcommand{\alpsym}{\ensuremath{\sim_\alpha}}
\newcommand{\choice}{\ensuremath{\chi}}
\newcommand{\ninb}{\ensuremath{\not\in_b}}
\newcommand{\inb}{\ensuremath{\in_b}}
\newcommand{\bvc}{\textbf{bvc}}

\newcommand{\noalp}{\ensuremath{no\alpha}}
\newcommand{\goes}[3]{\ensuremath{#1 : #2 \rightharpoonup #3}}
\newcommand{\goesd}[4]{\ensuremath{#1 : #2 \rightharpoonup #3 \downharpoonright #4}}

\newcommand{\restfresh}[3]{\ensuremath{#1 \mathbin{\#\!\!\downharpoonright} (#2,#3)}}

\newcommand{\pdash}{\raisebox{0.3em}{\ensuremath{\dashrightarrow}}\llap{\ensuremath{\dashrightarrow}}}
\newcommand{\pn}{\ensuremath{\rightrightarrows_n}}
\newcommand{\lc}{\ensuremath{\lambda}-calculus}
\newcommand{\lam}{\ensuremath{\lambda}}
\newcommand{\alp}{\ensuremath{\alpha}}
\newcommand{\alfa}{\ensuremath{\alpha}-conversion}
\newcommand{\alfac}{\ensuremath{\alpha}-convertible}
\newcommand{\alfaeq}{\ensuremath{\alpha}-equivalent}
\newcommand{\alfaeqe}{\ensuremath{\alpha}-equivalence}
\newcommand{\alfaeqc}{\ensuremath{\alpha}-equivalence classes}
\newcommand{\pa}{proof assistant}
\newcommand{\pas}{proof assistants}
\newcommand{\parallelsum}{\mathbin{\!/\mkern-5mu/\!}}
\newcommand{\pd}{\ensuremath{\dashrightarrow_\bet\!\!\!\!\!\!\!\!\!\!\parallelsum\ \ \ \ }}
\newcommand{\ra}{\ensuremath{\rightarrow}}
\newcommand{\alpCVesti}{\ensuremath{\stackrel{y\ \ \ }{\dashrightarrow_{i\alp^v}}}}
\newcommand{\alpCVest}{\ensuremath{\dashrightarrow_{\alp^v}}}
\newcommand{\alpCVestR}{\ensuremath{\dashrightarrow_{\alp^v}}}
\newcommand{\alpVest}{\ensuremath{\dashrightarrow_{\alp^v}^*}}
\newcommand{\betVest}{\ensuremath{\dashrightarrow_{\bet^v}}}
\newcommand{\capta}{\ensuremath{\text{Capt}}}
\newcommand{\capt}[2]{\ensuremath{\text{Capt}_{#1}(#2)}}
\newcommand{\eqdef}{\stackrel{\mathclap{\scriptsize\mbox{def}}}{=}}
\newcommand{\confa}{\ensuremath{\text{Conf}}}
\newcommand{\conf}[1]{\ensuremath{\text{Conf}(#1)}}
\newcommand{\fv}{\text{fv}}
\newcommand{\bv}{\text{bv}}
\newcommand{\alpeq}{\ensuremath{\alpha}-equivalence classes}
\newcommand{\Lam}{\ensuremath{\Lambda}}
\newcommand{\bet}{\ensuremath{\beta}}
\newcommand{\pen}{pen-and-paper}
\newcommand{\clc}[1]{\ensuremath{\mathbin{#1^{\mathcal{\lam}}}}}
\newcommand{\clrt}[1]{\ensuremath{\mathbin{#1^*}}}
\newcommand{\cl}[1]{\ensuremath{\mathbin{(#1^{\mathcal{C}})^*}}}
\newcommand{\eqH}[1]{\ensuremath{\lfloor #1 \rfloor_{H}}}
\newcommand{\eq}[2]{\ensuremath{\lfloor #1 \rfloor_{#2}}}
\newcommand{\eqR}[1]{\ensuremath{\lfloor #1 \rfloor}}
\newcommand{\lambdabar}{\mbox{\textipa{\textcrlambda}}}
\hyphenation{data-type}
\hyphenation{data-types}

\maketitle

\begin{abstract}
\vspace{-1em}
We introduce a universe of regular datatypes with variable binding information, for which we define generic formation and elimination (i.e. induction /recursion) operators. 
We then define a generic $\alpha$-equivalence relation over the types of the universe based on name-swapping, and derive iteration and induction principles which work modulo $\alpha$-conversion capturing  Barendregt's Variable Convention. We instantiate the resulting  framework so as to obtain the $\lambda$-Calculus and System F, for which we derive substitution operations and  substitution lemmas for $\alpha$-conversion and  substitution composition. 
The whole work is carried out in Constructive Type Theory and machine-checked by the system Agda.
\end{abstract}

\section{Introduction}

The definition of functions by recursion on the description of datatypes is the basic idea of generic programming. This method is based on defining a datatype, introduced as the \emph{universe}~\cite{mlof84:bibliopolis}, which contains datatype descriptions, such as ``a list is either empty or a pair consisting of a parameter and a sublist''. Indeed, the universe constructors correspond to the common notions ``either'', ``pair'', ``parameter'' and ``substructure'' abstracted out of informal descriptions such as the preceding one. Then a decoding function is introduced, which interprets instances of the universe, usually called universe \emph{codes}, into actual datatypes. 
In a dependently typed setting we can define generic functions over the universe of codes and the associated decoded datatypes. In other words, the codes provide information enough to properly traverse the structure of decoded datatype instances. Thereby, traversal becomes an operation that can be described for any recursive structure by means of generic iteration and recursion principles and, thus, code duplication is avoided by abstracting out basic operations on the datatypes. One can wonder how many other practical behaviors can undergo such kind of generalisation. 

In this work we introduce a  universe of \emph{regular trees}~\cite{Morris:2004} extended with variables (i.e. names) and binding information. We first define generic formation and elimination (i.e. induction/recursion) operators over this universe. The inclusion of names and the notion of locality allow us to introduce a generic \alp-equivalence relation, which we choose to base on name-swapping. Then we derive \alp-iteration and induction principles that capture Barendregt’s Variable Convention (BVC) which allows to proceed in proofs and definitions by conveniently choosing bound names so as to avoid conflict.
At this generic level we are able to prove several properties, mainly concerning the interaction of the iteration and recursion principles with the swapping operation and \alp-equivalence relation.
We next obtain \lc\ and System F as instances of our universe and define corresponding substitution operations as instances of the generic \alp-iteration principle.
We are thereby able to derive the lemmas on compatibility of substitution with \alp-equivalence by direct instantiation of the generic properties referred to above.
 Finally, we prove the substitution composition lemma for the System F showing how our approach allows us to mimic the BVC, proving particular  results on instances of the framework as it is usually done in pen-and-paper style.

\subsection{Related work} \label{sec:relatedwork}


%

Programming languages supporting native constructions to declare and manipulate abstract syntax with binders are presented by Shinwell, et. al in~\cite{ShinwellPG03,SHINWELL200653}, where an ML extension \emph{FreshML}, and an OCaml extension \emph{Fresh O'Caml} are respectively developed. These languages allow to deconstruct datatypes with binders in a safe way, that is, in the case of an abstraction inspection, a renaming with a freshly generated binder is computed for the abstraction body. In this way, the language user has access only to a fresh binder, and the renamed body of the opened abstraction. This mechanism grants that values with binders are operationally equivalent if they represent \alp-equivalent objects. This result is proved in~\cite{ShinwellPG03} by introducing a denotational semantics of the object language FreshML into FM-sets (Fraenkel and Mostowski's sets). They prove that this denotational semantics matches the operational one. In this way, they are able to prove that values of the introduced abstract syntax with binders  properly represent \alp-equivalence classes of the object-level syntax. In~\cite{Cheney} Cheney carries out a similar work, but instead of developing a language extension, he implements a Haskell library called \emph{FreshLib}. As the author does not implement a language from scratch, this work introduces generic programming techniques in its implementation  to support the required level of genericity.

All previous works address common operations dealing with general structures with binders. Although some of these developments give proofs about the soundness of their approaches, their main concern is the implementation of meta-programs.
In~\cite{Lee2012}, Lee et al.  use generic programming techniques to develop mechanisations of formal meta-theory in the Coq proof assistant.
This work allows the user to choose between nominal, locally nameless or de Bruijn first-order syntax.
For each of these representations, they offer several infrastructure operations and their associated lemmas.
For instance, for the locally nameless setting, two different substitutions are needed for bound and free variables respectively.
In the case of System F, where terms and type variables have binding constructions, this representation involves six different substitution operations.
Hence, as the number of syntactic sorts supporting binding constructions increases in the object language, there is a combinatorial explosion of the number of operations and lemmas involved in its formalisation for the locally nameless and de Bruijn first-order syntaxes.
They manage to address these issues defining these operations and associated lemmas in a generic re-usable way.
Moreover, they provide a small annotation language to describe the binding structure of the object language, from which they can automatically derive an isomorphism between the object language and their generic universe syntax.
However, introducing inductive relations in this framework requires the user to provide a mapping between the concrete relation,  defined at the object language level, and the generic relation.
They are able to instantiate some cases of the POPLmark challenge~\cite{Aydemir2005}  in their framework, validating their approach both for the locally nameless and the de Bruijn first-order syntax, and comparing some metrics of their approach against other solutions.
However, their particular choice of universe makes it impossible to have more than one sort of binder per datatype. Hence, they cannot represent in their setting a language such as Session Types~\cite{YOSHIDA200773}, where there exist three distinct sort of binders: parameters, channels and ports within a concurrent calculus.
We believe their work addresses reusing and usability in great manner, but lacks in extensibility and abstraction. By using this framework it is possible to reuse several operations and lemmas that hide some of the work required by the underlying  binders representation. However, in order to introduce new operations and prove results, the user may have to deal with the underlying generic abstract syntax language.
Their work seems to support the nominal syntax, although no \alp-conversion relation, neither any other classic relation, property or function over named terms is presented. Indeed, they do not further develop the nominal syntax, beyond the basic definitions of a nominal abstract syntax.

In~\cite{licata-harper-09}, Licata and Harper codify a universe that mixes binding and computation constructions in Agda, where computations are represented as meta-level functions injected in the universe constructions, i.e., they embed a HOL syntax in their development. Their representation is based on a well-scoped de Bruijn representation, that is, de Bruijn terms associated with a context indexing the free variables. For this universe, they provide a generic substitution operation, and prove context weakening and strengthening lemmas.
In 2002 Pitts and Gabbay introduced the \emph{Nominal Logic}~\cite{GP02:newapproach}, a first-order many-sorted logic with equality, containing primitives for renaming via name-swapping, for freshness of names, and for name-binding. The swapping operation has much nicer logical properties than the more general, non-bijective forms of renaming. 
This operation provides a sufficient foundation for a theory of structural induction/recursion for the syntax modulo \alp. In~\cite{UrbanT05}, Urban and Tasson use ideas from Nominal Logic to construct a set of \lam-calculus terms modulo alpha, that is, identifying \alp-convertible terms. The construction is based on a HOAS syntax on top of Isabelle/HOL, deriving recursion and induction principles over this quotient set.


Our main motivation is to show it is feasible to 
formalise
within constructive type theory  \alp-iteration/ induction principles for a classical named syntax, deriving these principles from just simple structural induction on fist-order terms, where equality remains the simple definitional one, and not performing any kind of quotient on terms. Then, we want to study how feasible is to use this development in practical examples.
This work is structured as follows: in Section~\ref{sec:rtrees} we present our regular tree universe, in Section~\ref{sec:nameswapping} we introduce name-swapping, in Section~\ref{sec:alpha} we give an \alp-conversion relation and iteration/induction schemes modulo \alp, which allow us to define the caputre-avoiding subtitution operation, and automatically derive some of its basic properties. In Section~\ref{sec:bvc} we introduce the proof technique that mimics the BVC, and we apply it in the substitution composition lemma. Finally, in the last section we discuss related work and conclusions.
We carry out the whole development within Constructive Type Theory as implemented in the system Agda~\cite{Norell2009}.
We will show fragments of the Agda code, the complete version being available at: \href{https://github.com/ernius/genericBindingFramework}{https://github.com/ernius/genericBindingFramework}.

\section{Universe of Regular Trees with Binders}
\label{sec:rtrees}
\subsection{Universe of (Codes of) Functors}
We choose to build up a universe whose objects are codes to be interpreted as functions from \AgdaDatatypeFoot{Set} to \AgdaDatatypeFoot{Set}\footnote{\AgdaDatatypeFoot{Set} is the type of (small) datatypes.}, i.e. as \emph{functors}. 
The actual datatypes generated by this mechanism arise as fixed points of such functors. 
To this effect, we introduce in Figure~\ref{fig:regulartree}:\begin{itemize}
    \item the datatype \AgdaDatatypeFoot{Functor} of codes, and
    \item the (mutually recursive-inductive) definition of the decoding function \AgdaFunctionFoot{⟦\_⟧} and of the actual datatype \AgdaDatatype{μ}\,$F$ associated to any given functor code $F$.
\end{itemize} 


\begin{figure}[ht]
\AgdaTarget{Functor}
\begin{minipage}[t]{0.45\linewidth}
\vspace{0.5em}
\small\ExecuteMetaData[GenericProgramming/GPBindings.tex]{functor}
\end{minipage}
\begin{minipage}[t]{0.45\linewidth}
  \small  \ExecuteMetaData[GenericProgramming/GPBindings.tex]{interpret}
  \AgdaTarget{μ}
  \small\ExecuteMetaData[GenericProgramming/GPBindings.tex]{mu}
\end{minipage}  
  \caption{Regular tree universe with binders.}
\vspace{-1em}
\label{fig:regulartree}
\end{figure}

Notice first that the (inductive) definition of \AgdaDatatype{μ}\,$F$ by means of the constructor $\langle\_\rangle$ indeed introduces it as the least fixed point of the functor corresponding to the code $F$. Now let us examine the codes and corresponding functors.
The first three constructors of datatype \AgdaDatatypeFoot{Functor} in Figure~\ref{fig:regulartree} represent the embedding of: the unity type, a recursive position, and an arbitrary (i.e. externally given) datatype. The fourth constructor embeds a datatype representable in our universe, while the next two constructors represent the sum and product of types. Finally, the last two introduction rules are specific to our desired domain of abstract syntaxes with binders. As our framework supports different sorts of names, the variables and binders constructors receive as parameters an identifier of the sort of variables that they respectively introduce or bind. The binder constructor also receives the descriptor of the structure serving as scope of the bound variable. In many cases of interest (e.g. the \lc\ and System F to be examined shortly) this subterm descriptor will be just a recursive position. However, a compound structure is often needed, as it is the case in e.g. languages with a \texttt{letrec} primitive.
We can observe how the variable and binder constructions inject a fixed set of variables (names) $V$\ into the interpreted datatype. This set $V$\ is assumed to be infinite with a decidable equality. 
Notice too that in the cases of the variable injection and the binder functors the sort argument $S$\ has no impact on the interpreted set. Indeed, we have only one kind of names $V$. The sort identifier will be relevant to implement generic operations related to binding issues, as shown in subsequent sections.

For example, the types of natural numbers and of lists of natural numbers can be defined  as follows:\\
\vspace{-1em}

\begin{minipage}[b]{0.4\linewidth}
  \qquad $\text{FNat} = |1|\ |{+}|\ \text{|R|}$
\end{minipage}
\begin{minipage}[b]{0.4\linewidth}
         $\text{FListNat} = |1|\ |{+}|\ (\text{|Ef|}\ \text{FNat})\ |{\times}|\ \text{|R|}$
\end{minipage}

\begin{minipage}[b]{0.4\linewidth}
  \qquad $\text{Nat} = \mu\ \text{FNat}$
\end{minipage}
\begin{minipage}[b]{0.4\linewidth}
  $\text{ListNat} = \mu\ \text{FListNat}$
\end{minipage}

\vspace{-1em}


\hfill

In Figure~\ref{fig:lcalc} we illustrate the use of the variables and binders constructions by encoding the $\lambda$-calculus. We show the corresponding classical concrete syntax  definition using comments, that are written following a dash to the right of each line. This definition has only one sort of variables identified with the sort \AgdaDatatypeFoot{SortλTermVars}. 

\begin{figure}[h!]
\begin{minipage}[t]{0.6\linewidth}
\small\ExecuteMetaData[GenericProgramming/Examples/LambdaCalculus.tex]{lambdacalculus}    
\end{minipage}
\begin{minipage}[t]{0.3\linewidth}
\small\ExecuteMetaData[GenericProgramming/Examples/LambdaCalculus.tex]{lambdacalculusmu}
\end{minipage}
\vspace{-1em}
\caption{\lam-calculus.}
\label{fig:lcalc}
\end{figure}

\vspace{-.5em}

We next introduce notation resembling the concrete syntax of the \lc\ and hiding away our universe code constructions.\\

\vspace{-1em}

\begin{minipage}{0.2\linewidth}
\small\ExecuteMetaData[GenericProgramming/Examples/LambdaCalculus.tex]{lambdaconstvar}  
\end{minipage}
\begin{minipage}{0.37\linewidth}
\small\ExecuteMetaData[GenericProgramming/Examples/LambdaCalculus.tex]{lambdaconstapp}
\end{minipage}  
\begin{minipage}{0.33\linewidth}
\small\ExecuteMetaData[GenericProgramming/Examples/LambdaCalculus.tex]{lambdaconstlam}
\end{minipage}

\hfill

Next we present the codification of the System F. As this language also needs bindings  at the type level, this encoding illustrates the use of two distinct sorts of identifiers, namely \AgdaDatatypeFoot{SortFTypeVars} and \AgdaDatatypeFoot{SortλTermVars}:

\begin{minipage}[t]{0.65\linewidth}  
\small\ExecuteMetaData[GenericProgramming/Examples/SystemF.tex]{systemF}
\end{minipage}
\begin{minipage}[t]{0.3\linewidth}
\small\ExecuteMetaData[GenericProgramming/Examples/SystemF.tex]{systemFmuty}
\\  
\\
\\
\small\ExecuteMetaData[GenericProgramming/Examples/SystemF.tex]{systemFmutrm}  
\end{minipage}

\hfill

In the preceding constructions we have
chosen a simplification of the universe of regular tree datatypes presented in~\cite{Morris:2004}, where recursive types are represented using $\mu$-types (from~\cite{Pierce:2002}). However, instead of the nominal approach traditionally used with recursive type binders, they use a well-scoped de Bruijn representation. Therefore, in order to properly interpret the full universe, a definition indexed by a context with the multiple $\mu$-recursive positions definitions  is required. Our representation in Agda (Figure~\ref{fig:regulartree}), simplifies this burden at the expense of not being able to represent mutually recursive datatypes. In other words, our universe construction has expressive power equivalent to admitting only a single top-level $\mu$-recursive type binder. 
\vspace{-1em}

\subsection{Map and Fold}
\label{sec:map-fold}

The classical definition of \emph{fold} based on \emph{map} that is usually introduced in category theory does not pass Agda's termination checker. The recursive call to fold is hidden inside a call to map, and because of this the termination checker cannot determine how map is using it. To make the fold operation pass the termination checker we have to fuse map and fold into a single function, as done in~\cite{Norell2009} for a similar regular tree universe. In  Figure~\ref{fig:foldt} we show our implementation of the function  \AgdaFunctionFoot{foldmap}. We make use of Agda's implicit arguments feature, denoted by curly braces, to omit terms that the type checker can figure out for itself. For instance, we declare the $A$ set argument as an implicit argument. The presented \AgdaFunctionFoot{foldmap} function needs to keep two functors, since the fold (recursive) part works always over the same functor argument $F$, while, for the map part, the auxiliary functor argument $G$\ gives the position of functor $F$ during the traversal of the structure. Therefore, this function only uses the functor $F$\ in the recursive case rule (the \AgdaInductiveConstructorFoot{|R|} case) in which the right hand side expression basically begins a new traversal of the functor $F$, in a way similar  to the original definition of fold. It does so by providing with a fresh copy of  $F$\ in the position of the auxiliary argument $G$. The rest of the rules are equivalent to a map over the functor $G$. Note that this definition terminates because the argument of type \AgdaDatatypeFoot{⟦ G ⟧ (μ F)} decreases in each recursive call. The new \AgdaFunctionFoot{fold} operation is defined as a recursive instance of \AgdaFunctionFoot{foldmap}.

\begin{figure}[h!]
\small\ExecuteMetaData[GenericProgramming/GPBindings.tex]{foldmap} \\
\small\ExecuteMetaData[GenericProgramming/GPBindings.tex]{foldmap2}
  \vspace{-2em}
  \caption{Terminating fold operation.}
\vspace{-.5em}
\label{fig:foldt}
\end{figure}

As an example, we define a function \AgdaFunctionFoot{vars} that counts the number of variable occurrences in a term of the $\lambda$-calculus. We do so by  instantiating it as a case of the  \AgdaFunctionFoot{fold} operation in fig.~\ref{fig:vars}



\begin{figure}[h!]
\begin{minipage}{0.45\linewidth}
\small\ExecuteMetaData[GenericProgramming/Examples/LambdaCalculus.tex]{varsfold1}
\end{minipage}
\begin{minipage}[b]{0.45\linewidth}
\small\ExecuteMetaData[GenericProgramming/Examples/LambdaCalculus.tex]{varsfold2}
\end{minipage}

     \caption{Fold application example.}
 \label{fig:vars}
  \vspace{-1em}
 \end{figure}

Next we present a particular useful instantiation of the fold operator, named \AgdaFunction{foldCtx}. This instantiation aims at reproducing some techniques related to the nominal syntax considered in our work. We introduce an extra argument with type \AgdaDatatypeFoot{μ C}, which is used by the folded function $f$. This function is partially applied to this extra argument, and then passed as an argument of fold. Hence, this argument acts as an explicit invariant context for the function $f$\ through the entire fold operation. Another difference with the original fold operation is that the result of this instance is a datatype \AgdaDatatypeFoot{μ H} encoded in our universe instead of an arbitrary set. \\
 \vspace{-1em}

{\small\ExecuteMetaData[GenericProgramming/GPBindings.tex]{foldCtx}}

\vspace{-1em}

From this fold instance we can directly derive the naive substitution operation for the \lc. In order to do this, we next give the functor descriptor \AgdaDatatypeFoot{cF} for the context argument. It represents the pair formed by the variable to be replaced and the substituted term:  \\

\vspace{-1em}

{\small\ExecuteMetaData[GenericProgramming/Examples/LambdaCalculus.tex]{substcontext}}

\vspace{-1em}

Next we define the function \AgdaFunctionFoot{substaux} (Figure~\ref{fig:substaux}) to be folded which,  given a term structure with the results of the recursive calls in its recursive positions, constructs the final result of the substitution. For the variable case, we check, as usual, whether the substitution is to be applied, whereas the application and abstraction cases directly reconstruct the  corresponding terms from the recursive call. Note that in the abstraction case we do not check whether the abstracted variable is different from the  one being replaced, as in Barendregt's substitution definition in~\cite{bar:84}. In fact, this comparison would be pointless because, as we are using an iteration principle, we do not have access to the original abstraction body subterm. Note that we hide the universe codes on the right side of this definition by using the previously introduced \lc\ constructors. 
\vspace{-.5em}

\begin{figure}[h!]
{\small\ExecuteMetaData[GenericProgramming/Examples/LambdaCalculus.tex]{substaux}}
  \vspace{-2em}
  \caption{Naive substitution auxiliary function.}
\vspace{-.5em}
\label{fig:substaux}
\end{figure}

Finally, we instantiate the \AgdaFunctionFoot{foldCtx} function with  \AgdaFunctionFoot{substaux}, and its appropriate context pair to get the naive substitution operation. \\

\vspace{-1em}

{\small\ExecuteMetaData[GenericProgramming/Examples/LambdaCalculus.tex]{naivesubst}}
\vspace{-2.45em}
\subsection{Primitive Induction}\label{sec:pimind}

We now develop a more generic elimination rule than the fold operation defined above. This elimination rule captures proof by induction, and is based on the recursion rule given by Benke et al. in~\cite{Benke:2003}. However, our development departs from their work in the following points: Firstly, they derive an elimination rule for a simpler universe construction, based on one-sorted term algebras, and defined through signatures instead of functors. For instance, their universe does not allow the injection of previously defined datatypes, which is necessary for defining e.g. lists of natural numbers so as natural numbers become parts of the lists. Secondly, their induction principle would not pass Agda's termination checker due to reasons similar to the ones discussed for the first version of the fold operation.
To define the desired function, to be called \AgdaFunctionFoot{foldInd}, we first introduce  the auxiliary function \AgdaFunctionFoot{fih} (Figure~\ref{fig:fih}).
\vspace{-.5em}

\begin{figure}[h!]
\small
  \ExecuteMetaData[GenericProgramming/GPBindings.tex]{primIndih}
\vspace{-2em}
  \caption{\AgdaFunctionFoot{fih} function.}
\vspace{-1em}
\label{fig:fih}
\end{figure}

This function receives a predicate $P$\ over the fixed point of a functor $F$, and an auxiliary functor $G$ (with a similar functionality to the one used in the \AgdaFunctionFoot{foldmap} function). It returns a corresponding predicate of type \AgdaFunctionFoot{⟦ G ⟧ (μ F) → Set}. This resulting predicate represents $P$\ holding for every recursive  position \AgdaDatatypeFoot{μ F} in an element of type \AgdaDatatypeFoot{⟦ G ⟧ (μ F)}.

We can now present our induction principle.  
 We proceed in a similar way as we did above for the fold function. First,  we introduce the fold-map fusion function \AgdaFunctionFoot{foldmapFh} (Figure~\ref{fig:ind}). Then,  we use this function to directly derive the induction principle as a recursive instance of the fold-map fusion. 

\begin{figure}[h!]
\small \ExecuteMetaData[GenericProgramming/GPBindings.tex]{primInd} \\
\small \ExecuteMetaData[GenericProgramming/GPBindings.tex]{primInd2}
\vspace{-2em}
 \caption{Induction principle.}
\vspace{-1em}
\label{fig:ind}
\end{figure}

We next give an example of the use of this induction principle, namely proving that the application of the function \AgdaFunctionFoot{vars} to any lambda term is greater than zero. We introduce the predicate \AgdaFunctionFoot{Pvars} representing the property to be proved and an auxiliary lemma \AgdaFunctionFoot{plus>0}, stating that the sum of two positive numbers is also positive. \\

\vspace{-1em}

\begin{minipage}[b]{0.3\linewidth}
\small\ExecuteMetaData[GenericProgramming/Examples/LambdaCalculus.tex]{varsproof1}
\end{minipage}
\begin{minipage}[b]{0.4\linewidth}
\small\ExecuteMetaData[GenericProgramming/Examples/LambdaCalculus.tex]{varsproof2}
\end{minipage}
\vspace{.5em}

The proof that \AgdaFunctionFoot{Pvars} holds for every term \AgdaFunctionFoot{M} is a direct application of the induction principle. 
The variable case is direct, while the application case is the application of the  lemma \AgdaFunctionFoot{plus>0} to the induction hypotheses. Finally, the abstraction case is a direct application of the induction hypothesis.
\vspace{.5em}

\begin{minipage}[t]{0.55\linewidth}
{\small\ExecuteMetaData[GenericProgramming/Examples/LambdaCalculus.tex]{proofpvars}}
\end{minipage}
\begin{minipage}[t]{0.3\linewidth}
{\small\ExecuteMetaData[GenericProgramming/Examples/LambdaCalculus.tex]{proof}}
\end{minipage}


\vspace{-2em}

\section{Name-Swapping}\label{sec:nameswapping}
\vspace{-1em}

We now turn to considering a very basic primitive of name-swapping, which will be used for defining 
$\alpha$-conversion without a mention to substitution. This  constitutes the foundation of the implementation of the general idea that principles of recursion and induction ought to be defined so as to work modulo 
$\alpha$-conversion, thus allowing to mimic the usual pen-an-paper conventions that allow the choice of convenient representatives of the terms involved in a definition or proof.
The name-swapping operation completely traverses a data structure, swapping occurrences (either free, bound or binding) of two given names of some sort.
Its implementation (Figure~\ref{fig:swap}) is similar to that of fold. 
\begin{figure}[h!]
 \small\ExecuteMetaData[GenericProgramming/Swap.tex]{swapF} \\
 \small\ExecuteMetaData[GenericProgramming/Swap.tex]{swap}
 \vspace{-2em}
  \caption{Name-swapping operation.}
\vspace{-1em}
 \label{fig:swap}
 \end{figure}

We use an auxiliary function \AgdaFunctionFoot{swapF}, that takes functors $F$ and $G$, and traverses the $G$ structure until a recursive or embedded position is reached, from where we restart the $G$ argument with either the original recursive functor $F$\ or the embedded functor respectively. Note that this treatment differs from the one in the definition of fold, where this case is a base case. Here we must also traverse the embedded functor instance, as we are swapping all the variables in the structure, including the variables present in any embedded structure. Because of this, we cannot derive name-swapping as an instance of fold. In the variable and abstraction cases we use name-swapping over variables~\footnote{Requiring the decidable equality over names.}, denoted by the (mixfix) operator \AgdaFunctionFoot{（_∙_）ₐ_} \ as in~\cite{CopelloTSBF16}. 

 We prove a generic lemma about the interaction between name-swapping and the iteration principle. This lemma is presented in Figure~\ref{fig:swapfoldCtx}, and states that the fold instance with context information is well-behaved with respect to name-swapping, given that the respectively folded operation is also well-behaved. Its proof  goes by a direct induction on terms. This example shows how we are able to develop generic proofs over our universe with binders. 

\begin{figure}[h!]
\small
  \ExecuteMetaData[GenericProgramming/Swap.tex]{swapfoldCtx}
 \vspace{-2em}
  \caption{Fold with context is well-behaved with respect to name-swapping.}
\vspace{-.5em}\label{fig:swapfoldCtx}
\end{figure}

We are able to directly apply the preceding lemma to the $\lambda$-calculus case in order to prove the result in Figure~\ref{fig:swapsubst}. 
This states that name-swapping commutes with substitution, which is  particularly useful. We introduce the  operator \AgdaFunctionFoot{（_∙_）_} \ to denote  the swapping of  variables in terms.
In the proof we use of the auxiliary lemma \AgdaFunctionFoot{lemma-substauxSwap} which states that the function \AgdaFunctionFoot{substaux}, used to define substitution, is well-behaved with respect to name-swapping. This example shows how feasible it is in our framework to instantiate generic proofs for deriving useful lemmas holding for particular instances of the generic universe. 

\begin{figure}[h!]
\small
  \ExecuteMetaData[GenericProgramming/Examples/LambdaCalculus.tex]{swapsubst}
\vspace{-2em}
  \caption{Substitution  is well-behaved with respect to name-swapping.}
\vspace{-1em}
\label{fig:swapsubst}
\end{figure}

In a similar manner we introduce a generic function returning the free variables of terms, and prove several properties about its interaction with swapping, fold, and \alp-conversion.

\vspace{-1em}

\section{Alpha Equivalence Relation.} \label{sec:alpha}
\vspace{-.5em}

\begin{figure}[h!]
\small
  \ExecuteMetaData[GenericProgramming/Alpha.tex]{alpha}
\vspace{-2em}
  \caption{Alpha equivalence relation.}
\vspace{-.5em}
\label{fig:alpha}
\end{figure}

In Figure~\ref{fig:alpha} we introduce the generic definition of the  \alp-equivalence relation over our universe, named \AgdaDatatypeFoot{∼α}. Its definition follows a  process similar to the one used before to implement generic functions over our universe. First, we define an auxiliary relation \AgdaDatatypeFoot{∼αF}, which is inductively defined introducing an auxiliary functor $G$, used to traverse the functor $F$\ structure. For the interesting binder case, we follow an idea similar  to the one used in~\cite{CopelloTSBF16}, that is, we define that two abstractions are \alp-equivalent if there exists some list of variables $xs$, such that for any given variable $z$\ not in $xs$, the result of swapping the corresponding binders with $z$\ in the abstraction bodies is \alp-equivalent. Note that the swapping is performed only over the sort of variables bound by this binder position, leaving any other sort of variables unchanged.
We are able to prove that this is an equivalence relation, and also that it is preserved under name-swapping in a similar way as done in our previous work~\cite{CopelloTSBF16}.

As we did before with name-swapping, we study how the iteration principle interacts with the introduced \alp-equivalence relation. We begin proving that the fold operation is \alp-compatible if it is applied to an also \alp-compatible function. We say a function is \alp-compatible iff it returns \alp-convertible results when it is applied to \alp-convertible arguments. In Figure~\ref{fig:foldalphaf} we state this lemma, whose proof goes by induction on terms. The only interesting case is the binders case, where we make use of the preservation of \alp-equivalence under name-swapping.

\begin{figure}[h!]
\small
  \ExecuteMetaData[GenericProgramming/Alpha.tex]{lemma-foldfalpha}
\vspace{-2em}
  \caption{Fold function \alp-compatibility property.}
\vspace{-1em}
\label{fig:foldalphaf}
\end{figure}

As a direct corollary we get that  the fold with context instance is \alp-compatible in its context argument provided the folded function is also \alp-compatible on its arguments (Figure~\ref{fig:foldalphafCtx}).\\

\begin{figure}[h!]
{\small
  \ExecuteMetaData[GenericProgramming/Alpha.tex]{lemmafoldCtxalphactx}}
\vspace{-2em}
 \caption{Fold context function \alp-compatibility corollary.} \label{fig:foldalphafCtx}
\vspace{-1em}
 \end{figure}

We define other relations over our universe in a similar way as we have done for the \alp-equivalence relation. For instance, the \AgdaDatatypeFoot{notOccurBind} relation 
holds if some given variable does not occur in any binder position within a term. In this relation 
we discard the name sort information. We do so to simplify our next development as we will explain later. We find useful to extend this relation to lists of variables, named as \AgdaDatatypeFoot{ListNotOccurBind}, which holds if all the variables in a given list do not occur in any binder position (associated with any sort) in a term. Using this relation we are able to prove the lemma stated in Figure~\ref{fig:foldalphacomp}. This lemma states that the fold with context principle is \alp-compatible on its two arguments if the provided function is \alp-compatible and well-behaved with respect to name-swapping. Note that this lemma extends the one given before in Figure~\ref{fig:foldalphafCtx}, although it requires extra freshness premises, and that the folded function is preserved under name-swapping.
\vspace{-.5em}

\begin{figure}[h!]
\small
  \ExecuteMetaData[GenericProgramming/Alpha.tex]{lemmafoldCtxalpha}
\vspace{-2em}
  \caption{Fold context \alp-compatibility property.}
\label{fig:foldalphacomp}
\end{figure}


\vspace{-2em}

\subsection{Alpha Fold} \label{sec:alphafold}
  
We are now able to introduce a fold operation that works at the level of \alp-equivalence classes of terms, that is, it only defines \alp-compatible functions. 
First, we introduce the function \AgdaFunctionFoot{bindersFreeElem} that takes a list of variables $xs$\ and an element $e$, and returns an element \alp-equivalent to $e$ whose binders are not in the given list. This function will be useful to reproduce the BVC, which basically states that we can always pick a term with its binders fresh from a given context, which in this function is represented as a list of variables. We prove that this function has the important property of being strongly \alp-compatible, i.e. that it returns the same result for \alp-convertible terms. \\

\vspace{-1em}

{\small\ExecuteMetaData[GenericProgramming/AlphaInduction.tex]{bindersfreealphaelem}} 

\vspace{-1em}

Based on this function, we next  directly implement the \alp-fold principle as an instance of the fold with context function. \\
\vspace{-1em}
  
{\small \ExecuteMetaData[GenericProgramming/AlphaInduction.tex]{foldCtxalpha} }
\vspace{-1em}


This iteration principle first finds a fresh term for a given context $c$, and then directly applies the fold operation over it. We developed this iteration principle following a different approach from the one taken in our previous work~\cite{CopelloTSBF16}, where we renamed the binders during the fold traverse. Instead, we chose to separate these two stages in order to reuse the previously defined fold operation and its properties. 

We can now properly justify the name ``alpha'' given to the introduced iteration principle. 
Firstly, as \AgdaFunctionFoot{bindersFreeElem} returns syntactical equal terms when applied to \alp-convertible terms, we have that our function is trivially strong \alp-compatible on its last term argument. 
Secondly, as a direct consequence from the lemma already proved for our iteration principle \AgdaFunctionFoot{foldCtx} in Figure~\ref{fig:foldalphafCtx}, this new principle inherits its \alp-compatibility in its context argument from \AgdaFunctionFoot{foldCtx}, given that the function received is also \alp-compatible on its arguments. Thus, the presented iteration principle works at the \alp-equivalence classes level when the given function works at the same level.
  


    


Now we can derive the  capture avoiding substitution operation for the lambda calculus example by  a direct application of the introduced \alp-fold principle. In fact this definition is exactly the same as the one given before for the naive substitution, but using now the \alp-fold operation instead of the fold one. \\
\vspace{-1em}

{\small\ExecuteMetaData[GenericProgramming/Examples/LambdaCalculus.tex]{subst} }
\vspace{-1em}


Substitution lemmas stating that  substitution is well-behaved with respect to \alp-conversion are  inherited from the \alp-compatibility of the iteration principle, only requiring the \alp-compatibility property of the \AgdaFunctionFoot{substaux} function. As this auxiliary function is not recursively defined, this proof is just a simple case analysis, while the proof involving the recursive data-type traversal is resolved at the generic level.


Next lemma in Figure~\ref{fig:foldfoldalpha} relates the presented \alp-fold principle with the previously defined one, giving sufficient conditions under which the two principles return \alp-convertible terms. First, the folded function must be \alp-compatible on its two arguments, and also well-behaved with respect to name-swapping. Secondly, we need a freshness premise stating that the free variables in the context do not occur bound in the applied term.
\vspace{-.5em}

\begin{figure}[h!]
\small\ExecuteMetaData[GenericProgramming/AlphaInduction.tex]{lemmafoldCtxalpha}
\vspace{-2em}
\caption{Fold and \alp-fold relations.}
\vspace{-.5em}
\label{fig:foldfoldalpha}
\end{figure}
We can instantiate this lemma to the $\lambda$-calculus to get sufficient conditions under which the two presented substitution operations are \alp-convertible. Its proof requires two lemmas about the \AgdaFunctionFoot{substaux} function, one stating that it is \alp-compatible, and the other one stating that it is well-behaved under name-swapping. Both lemmas were already used in previous proofs.


\vspace{-1em}

\subsection{Alpha Induction Principle}

In this section we generalise previous works~\cite{CopelloTSBF16,Copello:LSFA17}, developing an \alp-induction principle for \alp-compatible predicates. Our presentation introduces an explicit premise about the \alp-compatibility of the predicate being proved, which in general is not explicitly mentioned in informal developments, but is certainly required to ensure that the predicate in question is actually about \emph{abstract} terms, i.e. not dependent on the choice of bound names. Hence, to prove some property over any term, this principle requires the user to prove the property only for terms with fresh enough binders, i.e. distinct from the variables in some given context, as is usually done under the BVC.
We derive this principle following a procedure similar to the one used to infer the principle in Section~\ref{sec:pimind}. We show below the interesting recursive and binder cases, the remaining ones being equivalent to those presented in Figure~\ref{fig:fih}. We define an auxiliary function \AgdaFunctionFoot{fihalpha}, which transforms a given predicate $P$\ over a datatype \AgdaDatatypeFoot{μ F} to a predicate over the datatype \AgdaDatatypeFoot{⟦ G ⟧ (μ F)}. This predicate states that $P$\ holds for every recursive position \AgdaDatatypeFoot{μ F} in a datatype \AgdaDatatypeFoot{⟦ G ⟧ (μ F)}. Besides, it adds freshness premises with respect to some given variables list $xs$\ in the recursive and binder cases of its definition: In the binder case, it states that the binder is not in the given list $xs$, while, in the recursive case, it states that no variable in $xs$\ occurs in a binder position in the recursive subterm $e$. \\

\vspace{-1em}

{\small \ExecuteMetaData[GenericProgramming/AlphaInduction.tex]{alphainductionhypotheses}
\small \ExecuteMetaData[GenericProgramming/AlphaInduction.tex]{alphainductionhypothesescases} }

\vspace{-1em}

We state this principle in Figure~\ref{fig:alphaind}. Its proof is similar to the \alp-fold principle's proof. We firstly use the function \AgdaFunctionFoot{bindersFreeElem} (from Section~\ref{sec:alphafold}) over the parameter $e$\ and the freshness context $xs$\ to get an \alp-equivalent term $e'$\ with binders not occurring in the list $xs$. 
Then we apply the primitive induction principle (Figure~\ref{fig:ind}) over the fresh term $e'$\ to prove the following predicate $P'$:
\vspace{-.3em}
\begin{center}
    $P'(x) \equiv (\forall c \in xs \Rightarrow c\ notOccurrBind\ x) \Rightarrow P(x)$
\end{center}
Finally, we apply the proof of predicate $P'$\ to the term $e'$ and its freshness hypothesis to obtain that $P\ e'$\ must hold. 
Hence, as the predicate $P$\ is \alp-compatible, and $e$ \alpsym $e'$, we  get that $P\ e$\ should also hold.

\begin{figure}[h!]
\small \ExecuteMetaData[GenericProgramming/AlphaInduction.tex]{alphainductionprinciple}
\vspace{-2em}
  \caption{Alpha induction principle.}
\vspace{-.5em}
\label{fig:alphaind}
\end{figure}

The proof of $P'$\ is done using an auxiliary lemma which recursively reconstructs a proof of \AgdaFunctionFoot{fihalpha} $P\ xs\ e$\ given that \AgdaFunctionFoot{fih} $P\ xs\ e$\ holds and that the binders of $e$\ do not occur in the context $xs$. This proof is just a generalisation of the one already presented in~\cite{Copello:LSFA17} for an equivalent \alp-induction principle for \lc. In this previous work we were also able to prove the Church-Rosser theorem for the \lc\ using this equivalent induction principle. Therefore, we conjecture that following the same procedure we would be able to achieve the confluence of \bet-reduction result within our generic framework. However, we sketch another approach in next section to prove the substitution composition lemma, a crucial lemma in the confluence proof.
\vspace{-1em}

\section{Codification of a BVC proof technique.} \label{sec:bvc}

In Figure.~\ref{fig:alphaproof} we show a result that validates the BVC and usual practices in common pen-and-paper proofs within our generic framework. 
It states that for any \alp-compatible predicate $P$, we can prove $P\ e$ for any term $e$  by just proving it for terms whose binders are all different from their own free variables and from the variables in an arbitrary list $xs$. As in previous induction principle, this technique requires the \alp-compatibility of the predicate being proved.
\vspace{-.5em}

\begin{figure}[h!]
\small\ExecuteMetaData[GenericProgramming/AlphaInduction.tex]{alphaproof}
\vspace{-2.5em}
    \caption{BVC proof principle.}
\vspace{-.5em}
\label{fig:alphaproof}
\end{figure}

To prove $P\ e$ for arbitrary $e$, we proceed as follows: 
We first find a fresh enough term $e'$\ such that \ $e' \alpsym e$ \ using the function  \AgdaFunctionFoot{bindersFreeElem}. 
Then, we can use the hypothesis for the fresh term $e'$\ to derive that $P\ e'$ \ holds. 
Finally, $P\ e$ must also hold, as $P$ is \alp-compatible.
We do not show the code of the proof, since it is similar to others previously presented.


Next we illustrate the use of this result to prove the substitution composition lemma for  System F. First, we prove this lemma for the naive substitution operation. Next we introduce the property to be proved. Note that an extra freshness premise stating that $x$\ does not occur bound in the term $L$ is required, since we use the naive substitution. \\
\vspace{-1em}

{\small \ExecuteMetaData[GenericProgramming/Examples/SystemF.tex]{substnaivecompositionpredicate} 
}

\vspace{-1em}

The proof is done using the structural induction principle (Figure~\ref{fig:ind}).
We show below the interesting abstraction case:

{\small\ExecuteMetaData[GenericProgramming/Examples/SystemF.tex]{substncompositionabstractioncase}}
\vspace{-1em}

This equational proof is constructed following the usual pen-and-paper practice: 
First we  push the substitution  inside the abstraction. Then, by the induction hypothesis we know that the composition of substitutions in the abstraction bodies are \alp-convertible, and hence we are able to prove that the entire abstraction is \alp-convertible too, using the auxiliary lemma \AgdaFunctionFoot{lemma∼+B}. Finally, we push back the substitutions outside the abstraction to conclude the proof.

Now we prove the substitution composition lemma for the capture-avoiding substitution operation using the introduced \alp-proof technique. We begin by defining the functor describing a triple of terms \AgdaDatatypeFoot{TreeTermF}. Then, we introduce the predicate \AgdaDatatypeFoot{PSComp} over triples, stating the composition lemma  for the  substitution. 

{\small\ExecuteMetaData[GenericProgramming/Examples/SystemF.tex]{substcompositionpredicate} }

\vspace{-.5em}

We prove that \AgdaDatatypeFoot{PSComp} is \alp-compatible with respect to triples of terms 
 by a direct equational proof using basically the previous substitution lemmas.
In Figure~\ref{fig:substcompproof} we show the core of the proof. It uses the preceding substitution lemmas to replace the classical substitution operations with the naive ones. This can be done because we have  freshness premises stating that  in the introduced context of triples all binders are different from the free variables in the involved terms, and also from variables $x$ and $y$. 
Finally, we work in very much the same way as we did at the beginning to recover the classical substitutions from the naive ones. 
There are many auxiliary lemmas and boilerplate code concerning the freshness premises involved in the last proof which we do not show in this presentation. These are hidden inside auxiliary lemmas as: \AgdaFunctionFoot{y:fvL-NB-M[x≔N]ₙ} and \AgdaFunctionFoot{y:fvL-NB-N} occurring in the  proof. 
The first of these lemmas, for instance, proves that neither the variable $y$\ nor the free variable binding in $L$\ occur bound in $M[x≔N]_n$, which  is easy to  verify from the freshness premises. However, we believe further work is necessary to automatise some of these proofs, or even rewriting the freshness relations in order to alleviate its handling.

Finally, we can use the introduced \alp-proof principle with the previous proof obligations to finish the proof. Note how, by applying the \alp-proof technique to a triple of terms, we were able to get sufficient freshness premises to develop a proof similar in structure to pen-and-paper ones in a direct manner. This is possible because in our generic framework we can state the \alp-equivalence of any structure (triples in this case), and not just language terms. 
\vspace{-1ex}

\begin{figure}[h!]
  {\small \ExecuteMetaData[GenericProgramming/Examples/SystemF.tex]{substitutioncompositionproof}}\vspace{-2em}
\vspace{-1em}
  \caption{Proof of Substitution composition lemma.}
\vspace{-1em}
\label{fig:substcompproof}\end{figure}

\section{Conclusions}
\label{sec:concl}

 We  address the formalisation of a general first order named syntax with multi-sorted binders by applying a combination of generic programming and nominal techniques to derive fold operations,  name-swapping, the \alp-conversion relation, and \alp-induction/iteration principles for any language abstract syntax with binders. 
We derive the \lc\ and System F as instances of the introduced general framework. For these examples we derive both the naive and the capture-avoiding substitutions as direct instances of the corresponding fold and \alp-fold principles. We directly inherit the classical substitution lemmas for the \alp-conversion, and the good behavior of substitution under the name-swapping from fold properties already proved at the generic level. We prove a lemma stating sufficient conditions under which the fold and \alp-fold functions are \alp-equivalent. Therefore, as substitution operations are direct instances of these iteration principles, we get in an almost free manner a result about the relation between the naive and the capture-avoiding substitution operations for the \lc\ and System F. This  result is particularly useful in the last proof, which is conducted using the introduced \alp-proof technique, that enables us to mimic the BVC in a generic setup.

Our work uses generic programming techniques to develop the meta-theory of abstract syntax with binders in a general way as in related works. But we choose to maintain names for binders like as usually done in informal practice. On the other hand, contrary to the historical standpoint, following ideas in~\cite{GP02:newapproach}, we give \alp-conversion a more fundamental role than that of the definition of substitution. Indeed, we verify that the name-swapping is powerful enough to define a theory of structural induction/recursion modulo \alp\ in a general way.

We generalise the \alp-recursion/induction  principles developed in \cite{CopelloTSBF16,Copello:LSFA17}. In these previous works we renamed binders within the fold traversal. Instead, in this work we separate these stages, managing to reuse the fold operation and its properties. We also present an \alp-proof technique which is not based on an induction principle as in the previous works, and thus can be used to prove properties over relations or composite datatypes. This is the case in the proof of the substitution composition lemma, where it is used to obtain freshness premises over a triple of terms, which is possible because of the generic character of the approach, allowing us to state the \alp-equivalence or freshness premises over any composite datatype; that is, we are able to state freshness in any mathematical context, thus reflecting the BVC more accurately.

Generic programming techniques are capable of further improvements as the one considered in~\cite{Delaware:2013:MLC}, where a more modular assembly is introduced, enabling a more structured approach to the reuse of meta-theory formalisations through the composition of modular inductive definitions and proofs. The present work does not directly support a modular reuse, but it would be interesting to explore this improvement.

In~\cite{DBLP:journals/corr/AnandM17} Reynold's parametricity theory is used to prove the \alp-compatibility property of a big step semantics using reflection within Coq. They introduce a lambda calculus terms interface, and by a formalisation of Reynold's parametricity, they prove that polymorphic functions (over this interface) applied to related inputs produces related outputs. Then, given two concrete implementations of their lambda terms interface, one with de Brujin syntax (where α-convertible are syntaticaly equal), and another one nominal, they are able to get as a "free theorem" that on α-convertible inputs, the big step function produces α-convertible outputs in the nominal syntax. It remains as future work to study how our generic framework could be used to internalise this kind of "free theorems" by introducing a de Brujin interpretation to our universe, and then translate results between interpretations. \\

\noindent \textbf{Acknowledgements.} We gratefully acknowledge DoD support under award FA9550-16-1-0082 (MURI Program).

\vspace{-1em}

\bibliographystyle{eptcs}
\bibliography{bibliography}
\end{document}